\documentclass[]{aastex631}
\newcommand{\rcom}{\textcolor{black}}
\newcommand{\calviacat}{{\sc calviacat }}
\newcommand{\degree}{^\circ}

\shorttitle{Analysis of 2023 RN$_{3}$ Cometary Activity}
\shortauthors{Dobson et al.}
\graphicspath{{./}{figures/}}

\begin{document}

\title{Analysing the Onset of Cometary Activity by the Jupiter-Family Comet 2023 RN$_{3}$}

\correspondingauthor{Matthew M. Dobson}
\email{mdobson04@qub.ac.uk}

\author[0000-0002-1105-7980]{Matthew M. Dobson}
\affiliation{Astrophysics Research Centre, School of Mathematics and Physics, Queen's University Belfast, Belfast BT7 1NN, UK}

\author[0000-0003-4365-1455]{Megan E. Schwamb}
\affiliation{Astrophysics Research Centre, School of Mathematics and Physics, Queen's University Belfast, Belfast BT7 1NN, UK}

\author[0000-0003-0250-9911]{Alan Fitzsimmons}
\affiliation{Astrophysics Research Centre, School of Mathematics and Physics, Queen's University Belfast, Belfast BT7 1NN, UK}

\author[0000-0002-6702-7676]{Michael S. P. Kelley}
\affiliation{Department of Astronomy, University of Maryland, College Park, MD 20742-0001, USA}

\author[0000-0002-4043-6445]{Carrie E. Holt}
\affiliation{Las Cumbres Observatory, 6740 Cortona Drive Suite 102, Goleta, CA 93117, USA}

\author[0000-0001-9505-1131]{Joseph Murtagh}
\affiliation{Astrophysics Research Centre, School of Mathematics and Physics, Queen's University Belfast, Belfast BT7 1NN, UK}

\author[0000-0002-4043-6445]{Henry H. Hsieh}
\affiliation{Planetary Science Institute, 1700 East Fort Lowell Rd., Suite 106, Tucson, AZ 85719, USA}
\affiliation{Institute of Astronomy and Astrophysics, Academia Sinica, P.O. Box 23-141, Taipei 10617, Taiwan}

\author[0000-0002-7034-148X]{Larry Denneau}
\affiliation{Institute for Astronomy, University of Hawaii at Manoa, Honolulu, HI 96822, USA}

\author[0000-0002-9986-3898]{Nicolas Erasmus}
\rcom{\affiliation{Department of Physics, Stellenbosch University, Stellenbosch, 7602, South Africa}}
\affiliation{South African Astronomical Observatory, Cape Town 7925, South Africa}

\author[0000-0003-3313-4921]{A. N. Heinze}
\affiliation{Department of Astronomy and the DIRAC Institute, University of Washington, 3910 15th Ave NE, Seattle, WA 98195, USA}

\author[0000-0002-5738-1612]{Luke J. Shingles}
\affiliation{GSI Helmholtzzentrum f\"{u}r Schwerionenforschung, Planckstraße 1, 64291 Darmstadt, Germany}
\affiliation{Astrophysics Research Centre, School of Mathematics and Physics, Queen's University Belfast, Belfast BT7 1NN, UK}

\author[0000-0001-5016-3359]{Robert J. Siverd}
\affiliation{Institute for Astronomy, University of Hawaii at Manoa, Honolulu, HI 96822, USA}

\author[0000-0001-9535-3199]{Ken W. Smith}
\affiliation{Astrophysics Research Centre, School of Mathematics and Physics, Queen's University Belfast, Belfast BT7 1NN, UK}

\author[0000-0003-2858-9657]{John L. Tonry}
\affiliation{Institute for Astronomy, University of Hawaii at Manoa, Honolulu, HI 96822, USA}

\author[0000-0003-1847-9008]{Henry Weiland}
\affiliation{Institute for Astronomy, University of Hawaii at Manoa, Honolulu, HI 96822, USA}

\author[0000-0002-1229-2499]{David. R. Young}
\affiliation{Astrophysics Research Centre, School of Mathematics and Physics, Queen's University Belfast, Belfast BT7 1NN, UK}

\author[0000-0002-3818-7769]{Tim Lister}
\affiliation{Las Cumbres Observatory, 6740 Cortona Drive Suite 102, Goleta, CA 93117, USA}

\author[0000-0001-5749-1507]{Edward Gomez}
\affil{Las Cumbres Observatory, School of Physics and Astronomy, Cardiff University, Queens Buildings, The Parade, Cardiff CF24 3AA, UK}

\author[0000-0002-1278-5998]{Joey Chatelain}
\affil{Las Cumbres Observatory, 6740 Cortona Drive Suite 102, Goleta, CA 93117, USA}

\author[0000-0002-4439-1539 ]{Sarah Greenstreet}
\affil{Rubin Observatory/NSF NOIRLab, 950 N. Cherry Ave, Tucson, AZ 85719, USA}
\affil{Department of Astronomy and the DIRAC Institute, University of Washington, 3910 15th Ave NE, Seattle, WA 98195, USA}

\begin{abstract}

We utilize serendipitous observations from the Asteroid Terrestrial-impact Last Alert System (ATLAS) and the Zwicky Transient Facility (ZTF) in addition to targeted follow-up observations from the Las Cumbres Observatory (LCO) and Liverpool Telescope to analyze the first observed instance of cometary activity by the newly-discovered Jupiter-family comet C/2023 RN$_{3}$ (ATLAS), whose orbital dynamics place it close to residing on a Centaur-like orbit. Across our \rcom{7}-month baseline, we observe an epoch of cometary activity commencing in August 2023 with an increase in brightness of ${>}5.4$ mag. The lightcurve of 2023 RN$_{3}$ indicates the presence of continuous cometary activity across our observations, suggesting the onset of a new period of sustained activity. We find no evidence of any outbursts on top of the observed brightening, nor do we find any significant color evolution across our observations. 2023 RN$_{3}$ is visibly extended in LCO and Liverpool Telescope observations, indicating the presence of a spatially-extended coma. Numerical integration of 2023 RN$_{3}$'s orbit reveals the comet to have recently undergone a slight increase in semimajor axis due to a planetary encounter with Jupiter, however whether this orbital change could trigger 2023 RN$_{3}$'s cometary activity is unclear. Our estimate for the maximum dust production metric of $Af\rho \sim$400 cm is consistent with previous measurements for the Jupiter-family comet and Centaur populations. 

\end{abstract}

\section{Introduction} \label{sec:intro}

C/2023 RN$_{3}$ (ATLAS) (hereafter refered to as ``2023 RN$_{3}$'') is a Jupiter-family comet (JFC) recently discovered by the Asteroid Terrestrial-impact Last Alert System (ATLAS; \citealt[]{2018PASP..130f4505T,2018ApJ...867..105T}) wide-field sky survey on 2023 September 4 \citep{Green2023} during an epoch of cometary activity which caused it to brighten by more than $4$ magnitudes \citep{2023RNAAS...7..263H}. 
Early analysis of 2023 RN$_{3}$'s cometary activity by \citet{2023RNAAS...7..263H} revealed the presence of a visible coma.  With a semimajor axis $a = 10.147$ au, perihelion $q = 5.172$ au (last passed on 2023 January 16), eccentricity $e = 0.490$, and inclination $i = 10.358$ deg, 2023 RN$_{3}$'s orbit is that of a JFC, with a Tisserand parameter $T_{J} = 2.907$. However, its perihelion and Tisserand parameter values lie close to the threshold values ($q = 5.2$ au \citep{2009AJ....137.4296J,2008ssbn.book...43G} and $T_{J} = 3.05$ \citep{2008ssbn.book...43G}, respectively) that separate the JFC population from the Centaurs, small icy objects on chaotic orbits in the giant planet region of the solar system that are thought to replenish the JFC population, making the study of 2023 RN$_{3}$ relevant to studies of both JFCs and Centaurs. 

Both Centaurs and JFCs are known to exhibit cometary activity \citep[]{2009AJ....137.4296J,2009P&SS...57.1133K,2013ApJ...773...22B,2020tnss.book..307P,2022arXiv220301397J}, manifesting as dust comae and tails. This activity can take the form of `outbursts', sudden and  brief increases in the mass-loss and corresponding brightness of the object \citep{1990QJRAS..31...69H}. However, while the activity of Jupiter-family comets is closely coupled with their proximity to perihelion, the activity exhibited by Centaurs can occur throughout their orbits \citep{2020tnss.book..307P}. This points to different mechanisms responsible for activity exhibited by both populations. The activity of Jupiter-family comets is generally due to the sublimation of water ice for those with perihelia interior to ${\sim}3$ au \citep{2004come.book..317M,2017PASP..129c1001W}, with sublimation of CO and CO$_2$ thought to be the dominant driver of JFC activity exterior to this distance \citep[]{2012ApJ...752...15O,2015SSRv..197....9C}. By comparison, the mechanism responsible for Centaur activity is not as well understood. Proposed causes of Centaur activity include newly-exposed pockets of volatile surface ices \citep{1992ApJ...388..196P} or volatiles released by the amorphous-to-crystalline transition of sub-surface water ice \citep{2009AJ....137.4296J}. Furthermore, analysis of Centaur orbital dynamics points to correlations between the activity exhibited by a Centaur and its perihelion \citep[]{2009AJ....137.4296J,2021PSJ.....2..155L}, its lifetime in the giant planet region \citep{2018P&SS..158....6F}, and its evolutionary history, with active Centaurs tending to have encountered a giant planet in the last ${\sim}10^{3}$ years of its history, causing an abrupt decrease in its semimajor axis \citep[]{2018P&SS..158....6F,2021PSJ.....2..155L,2024ApJ...960L...8L}. Understanding the mechanism behind Centaur activity, and how this activity transitions to that exhibited by Jupiter-family comets, is thus an important step in understanding cometary evolution in the solar system.

In this paper, we analyze the 2023 cometary activity of 2023 RN$_{3}$, combining follow-up observations from Las Cumbres Observatory (LCO, \citealt{2013PASP..125.1031B}) as part of the Las Cumbres Observatory Outbursting Objects Key (LOOK; \citealt{2022PSJ.....3..173L}) project and the Liverpool Telescope \citep{2004SPIE.5489..679S} along with serendipitous observations from ATLAS and the Zwicky Transient Facility (ZTF; \citealt[]{2019PASP..131a8002B,2019PASP..131g8001G}). This paper is structured as follows: in Section \ref{ObservationsAndDataReduction}, we briefly describe the observations used in our analysis and photometry. In Section \ref{AnalysisAndResults} we present our analysis and findings studying the evolution in the brightness, color, radial profile, and dust production of 2023 RN$_{3}$. We present our summary and conclusions in Section \ref{Discussion}.

\section{Observations and Data Reduction} \label{ObservationsAndDataReduction}

Our dataset consists of both targeted follow-up observations of 2023 RN$_{3}$ (LOOK and Liverpool Telescope) and serendipitous observations of the object from automated surveys (ATLAS and ZTF). Our observations, summarised in Table \ref{2023RN3SurveyDatasets}, extend in time from 2023 June 6 to 2024 February 2. The serendipitous observations from ATLAS and ZTF make up the majority of our dataset, and cover the onset of 2023 RN$_{3}$'s cometary activity first observed on 2023 August 26 (see Section \ref{BrightnessEvolution}). Our observations from LOOK and Liverpool telescope sample 2023 RN$_{3}$ after its activity commenced.

\begin{deluxetable*}{lcccr}[h] \label{2023RN3SurveyDatasets}
\tablecaption{Number and Time Ranges of Observations of 2023 RN$_{3}$ for Each Filter Used in our Analysis}
\tablecolumns{4}
\tablehead{
\colhead{Telescope/Survey} & 
\colhead{Filter} & 
\colhead{Start Date} & 
\colhead{End Date} &
\colhead{Number of Observations}\\
\colhead{} & 
\colhead{} & 
\colhead{(YYYY-MM-DD)} &
\colhead{(YYYY-MM-DD)} &
\colhead{}
}
\startdata
ATLAS&$c$&2023-07-18&2023-12-10&36\\
ATLAS&$o$&2023-06-06&2023-12-16&106\\
ZTF&ZTF-$g$&2023-08-03&2023-11-13&26\\
ZTF&ZTF-$r$&2023-08-03&2024-01-05&28\\
Liverpool Telescope&SDSS-$g'$&2023-12-08&2023-12-14&6\\
Liverpool Telescope&SDSS-$r'$&2023-11-26&2023-12-14&11\\
Liverpool Telescope&SDSS-$i'$&2023-11-26&2023-12-14&12\\
LOOK&PS1 $w$&2023-11-17&2023-11-17&4\\
LOOK&SDSS-$g'$&2023-11-16&2024-01-10&26\\
LOOK&SDSS-$r'$&2023-11-16&2024-01-10&24\\
LOOK&SDSS-$i'$&2023-11-26&2024-01-10&19\\
\hline
Total&&2023-06-06&2024-01-10&298\\
\enddata
\end{deluxetable*}

\subsection{ATLAS}

ATLAS regularly observes the entire night sky to a limiting magnitude of 19.5 mag using 30-second exposures in two non-standard filters, cyan ($c$, 420-650 nm) and orange ($o$, 560-820 nm) \citep{2018PASP..130f4505T}. 
ATLAS consists of four telescope units, two of which are located in Hawai'i (Mauna Load and Haleakala), one in Chile (El Sauce) and one in South Africa (Sutherland). Each of the four ATLAS units consists of a 0.5-m Schmidt telescope with a field of view (FOV) of 28.9 square degrees  with a pixel scale of 1.86 arcsec/pixel \citep[]{2018PASP..130f4505T,2018ApJ...867..105T}. 
Further details of ATLAS and its data reduction pipeline can be found in \citet{2018PASP..130f4505T,2018ApJ...867..105T}, and \citet{2020PASP..132h5002S}.
We use both $c$ and $o$ filter observations of 2023 RN$_{3}$ from all four ATLAS telescopes between 2023 June 6 and 2024 February 2. 
We utilize the difference images automatically generated by the ATLAS reduction pipeline for each observation, to reduce the effect of contamination from background flux sources (e.g. stars, galaxies) on our photometry of 2023 RN$_{3}$.

\subsection{ZTF}

Using the 48-inch (1.2-m) Samuel Oshin telescope, the Zwicky Transient Facility (ZTF) is a wide-field time domain survey which has been observing the visible sky since 2017 October with a cadence of approximately three days. Each observation has an exposure time of 30-seconds, with a FOV of 43 square degrees and a pixel scale of 1.01 arcsec/pixel, and are obtained using three broadband filters ZTF-$g$, ZTF-$r$, and ZTF-$i$. Further details of ZTF, its system overview, survey strategy, data processing pipeline, and science objectives can be found in \citet{2019PASP..131a8002B},  \citet{2019PASP..131g8001G}, and \citet{2019PASP..131a8003M}. We query the ZTF observations of 2023 RN$_{3}$ via the Infrared Science Archive (IRSA) of the National Aeronautics and Space Administration (NASA) Infrared Processing and Analysis Center (IPAC), which utilizes the Jet Propulaion Laboratory (JPL) Horizons\footnote{\url{https://ssd.jpl.nasa.gov}} ephemeris of a queried object to obtain all ZTF observations which coincide with the position of that object. These observations range in time from 2023 August 3 to 2024 January 7 in the broadband filters ZTF-$g$ and ZTF-$r$.

\subsection{LOOK}

The Las Cumbres Observatory (LCO) 1.0-m telescopes \citep{2013PASP..125.1031B} observed 2023 RN$_{3}$ from 2023 November 26 to 2024 January 10 (including the observations reported in \citealt{2023RNAAS...7..263H}) as part of the LCO Outbursting Objects Key Project (LOOK; \citealt{2022PSJ.....3..173L}). Observations were taken in SDSS $g$, $r$, and $i$ filters \citep{2016arXiv161205560C} with the original \citet{2023RNAAS...7..263H} observations in the $w$-filter included in our analysis. Observations were scheduled with the NEOExchange web portal \citep{2021Icar..36414387L}. Each 1-m telescope utilizes an identical Sinistro imager comprising a Fairchild $4096 \times 4096$ pixel charge-coupled device CCD with a field of view of $26.6' \times 26.6'$, resulting in a pixel scale of 0.387 arcsec/pixel \citep{2013PASP..125.1031B}. 
To ensure negligible on-sky motion of 2023 RN$_{3}$ within this pixel scale, all observations were obtained with the telescope in half-rate tracking mode, with exposure times of 245 seconds. These observations are automatically processed using the LCO image processing pipeline ``Beautiful Algorithm to Normalize Zillions of Astronomical Images" (BANZAI; \citealt{2018SPIE10707E..0KM}). The LOOK photometry pipeline includes automatic calibration of observations to the Pan-STARRS1 photometric system \citep{2012ApJ...750...99T} using the ATLAS-RefCat2 all-sky photometric catalog \citep{2018ApJ...867..105T}, the \calviacat software \citep{2021zndo...5061298K}, and background field stars measured with BANZAI \citep{2022PSJ.....3..173L}. Due to the difference between the Pan-STARRS1 filter system of ATLAS-RefCat2 and the SDSS filter system of LCO, the LOOK photometry pipeline performs a color correction to transform the instrumental magnitudes to their corresponding Pan-STARRS1 filters $g'$, $r'$, and $i'$ \citep{2022PSJ.....3..173L}. Of these observations, we utilize for our analysis only those which were obtained with seeing $\leq4$ arcsec.

\subsection{Liverpool Telescope}

The Liverpool Telescope is a 2-m robotic telescope located at the Observatorio del Roque de los Muchachos, La Palma, Canary Islands \citep{2004SPIE.5489..679S}.
We obtained observations of 2023 RN$_{3}$ between 2023 November 26 and 2023 December 14 using the IO:O wide-field camera \citep{2016JATIS...2a5002B} whose $4096 \times 4112$ pixel CCD results in a field of view $10' \times 10'$. Observations were obtained in SDSS $r$ and $i$ filters for conditions of airmass $<2.0$ and seeing $<2.0$ arcsec. We utilize sidereal tracking of 2023 RN$_{3}$, with exposure times of 82 seconds for $g$ filter observations and 70 seconds in $r$ and $i$ filters to ensure negligible on-sky motion of the object within the 0.3 arcsec/pixel resolution afforded by 2x2 pixel binning. Observations were automatically reduced via the IO:O reduction pipeline. We analyze 2023 RN$_{3}$ in the 29 observations where it was detected above $3\sigma$.

\section{Analysis and Results} \label{AnalysisAndResults}

Our multi-wavelength observations from ATLAS, ZTF, LOOK, and Liverpool Telescope allow us to analyze 2023 RN$_{3}$'s 2023 cometary activity. We use the multi-month, multi-wavelength baseline observations from ATLAS, ZTF, and LOOK to explore the evolution of 2023 RN$_{3}$'s measured brightness and determine any evolution of its observed color. We use the high-resolution LOOK and Liverpool Telescope observations to analyze 2023 RN$_{3}$'s surface brightness distribution across time. Finally, we utilize our ATLAS, ZTF, and LOOK observations to estimate 2023 RN$_{3}$'s activity level across our observational baseline.

\subsection{Brightness Evolution} \label{BrightnessEvolution}

Using our largest datasets, ATLAS, ZTF, and LOOK, we examine the brightness evolution of 2023 RN$_{3}$. To account for the visible extension exhibited by 2023 RN$_{3}$ as reported by \citet{2023RNAAS...7..263H}, we perform aperture photometry on all our observations with a sufficiently large aperture, applying an effective radius of 20,000 km, corresponding to an angular size of 4.83$\arcsec$ to 6.36$\arcsec$ across our observations. We use the aperture photometry automatically calculated by the LOOK photometry pipeline \citep{2022PSJ.....3..173L}.
For our ATLAS and ZTF observations, we use Source Extractor Python \citep[]{1996A&AS..117..393B,2016JOSS....1...58B} to identify sources. We identify the closest source as 2023 RN$_{3}$ within a threshold distance (3 arcsec for ATLAS observations, 1.5 arcsec for ZTF, due to the differing pixel scales of both surveys) from  2023 RN$_{3}$'s predicted position by JPL Horizons. We use the \textit{photutils} Python library \citep{larry_bradley_2023_7946442} to perform aperture photometry on 2023 RN$_{3}$, applying a circular aperture. The background annulus apertures had inner and outer radii of 4 and 5 times the circular aperture radius, 20,000 km, respectively. We apply per observation the zeropoints reported from the ATLAS \citep[]{2018ApJ...867..105T} and ZTF \citep{2019PASP..131a8003M} photometry pipelines. If the flux measured for 2023 RN$_{3}$ has a signal-to-noise ratio (SNR) of $< 3$, or if no source is found by Source Extractor Python, we calculate a $3\sigma$ upper limit $m_{3\sigma}$ from forced photometry at the predicted coordinates of 2023 RN$_{3}$ according to the equation:
\begin{equation}
    M_{3\sigma} = -2.5\log_{10}(3*dF) + ZP
\end{equation}
where $dF$ is the uncertainty in the measured flux and $ZP$ is the observation zeropoint magnitude. Table \ref{2023RN3Data} lists the magnitude measurements (or $3\sigma$ upper limits) of 2023 RN$_{3}$ for each observation on which we performed photometry.

\begin{deluxetable*}{lcccccccr}[h] \label{2023RN3Data}
\tablecaption{Apparent Magnitudes of 2023 RN$_{3}$ in Each Observation from ATLAS, ZTF, and LOOK.}
\tablecolumns{4}
\tablehead{
\colhead{MJD} & 
\colhead{Magnitude} & 
\colhead{Magnitude} &
\colhead{Photometry} & 
\colhead{Filter} &
\colhead{Heliocentric} & 
\colhead{Geocentric} & 
\colhead{Phase Angle} &
\colhead{Survey}\\
\colhead{} & 
\colhead{} & 
\colhead{Uncertainty} &
\colhead{Flag} &
\colhead{} &
\colhead{Distance} &
\colhead{(au)} &
\colhead{Distance} &
\colhead{}\\
\colhead{} & 
\colhead{(mag)} & 
\colhead{(mag)} &
\colhead{} &
\colhead{} &
\colhead{(au)} &
\colhead{(au)} &
\colhead{(deg)} &
\colhead{}\\
}
\startdata
60191.567824&17.683&0.084&1&o&5.312&4.377&4.5017&ATLAS\\
60191.571448&17.875&0.100&1&o&5.312&4.377&4.5011&ATLAS\\
60191.604028&17.783&0.099&1&o&5.312&4.377&4.4952&ATLAS\\
60204.173872&17.677&0.065&1&o&5.327&4.339&2.1776&ATLAS\\
60204.177039&17.635&0.062&1&o&5.327&4.339&2.1770&ATLAS\\
60204.179792&17.821&0.087&1&o&5.327&4.339&2.1765&ATLAS\\
60204.210026&17.660&0.071&1&o&5.327&4.339&2.1711&ATLAS\\
60213.104433&17.628&0.108&1&o&5.339&4.339&0.9948&ATLAS\\
60213.127136&17.588&0.101&1&o&5.339&4.339&0.9942&ATLAS\\
60213.140634&17.678&0.108&1&o&5.339&4.339&0.9939&ATLAS\\
\enddata
\tablecomments{This table is published in its entirety in the machine-readable format. A portion is shown here for guidance for regarding its form and content.}
\tablecomments{Flag indicates if brightness measurement is the measured magnitude (flag = 1) or a $3\sigma$ upper limit (flag = 0).}
\end{deluxetable*}

\begin{figure}
\centering
\includegraphics[width=\textwidth]{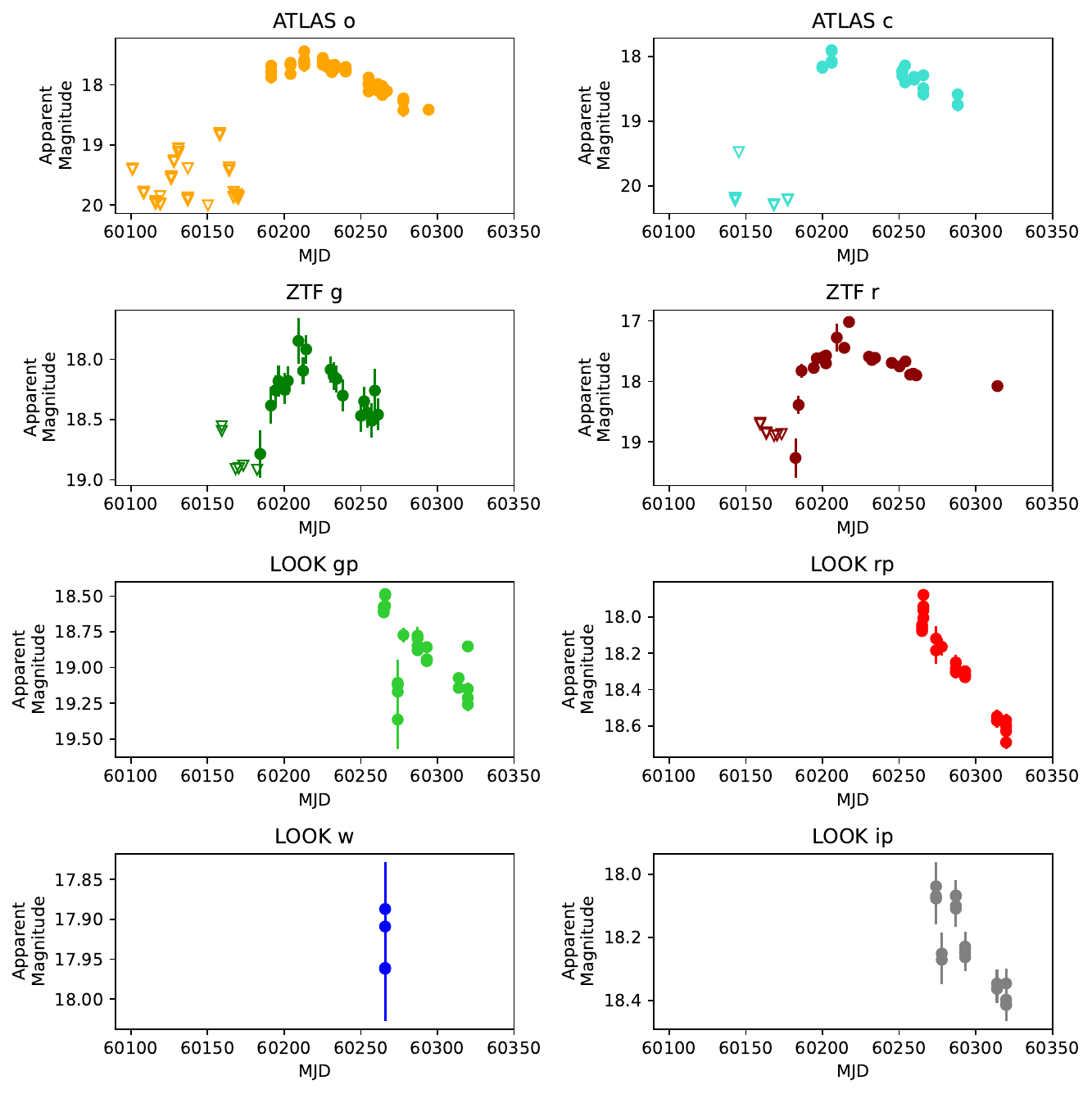}
\caption{Lightcurve of 2023 RN$_{3}$ (apparent magnitude across time) in each filter. Filled circles indicate magnitude measurements, empty markers triangle indicate $3\sigma$ upper limits.}
\label{2023RN3ApparentMagnitudeVsTime}
\end{figure}

\begin{figure}
\centering
\includegraphics[width=\textwidth]{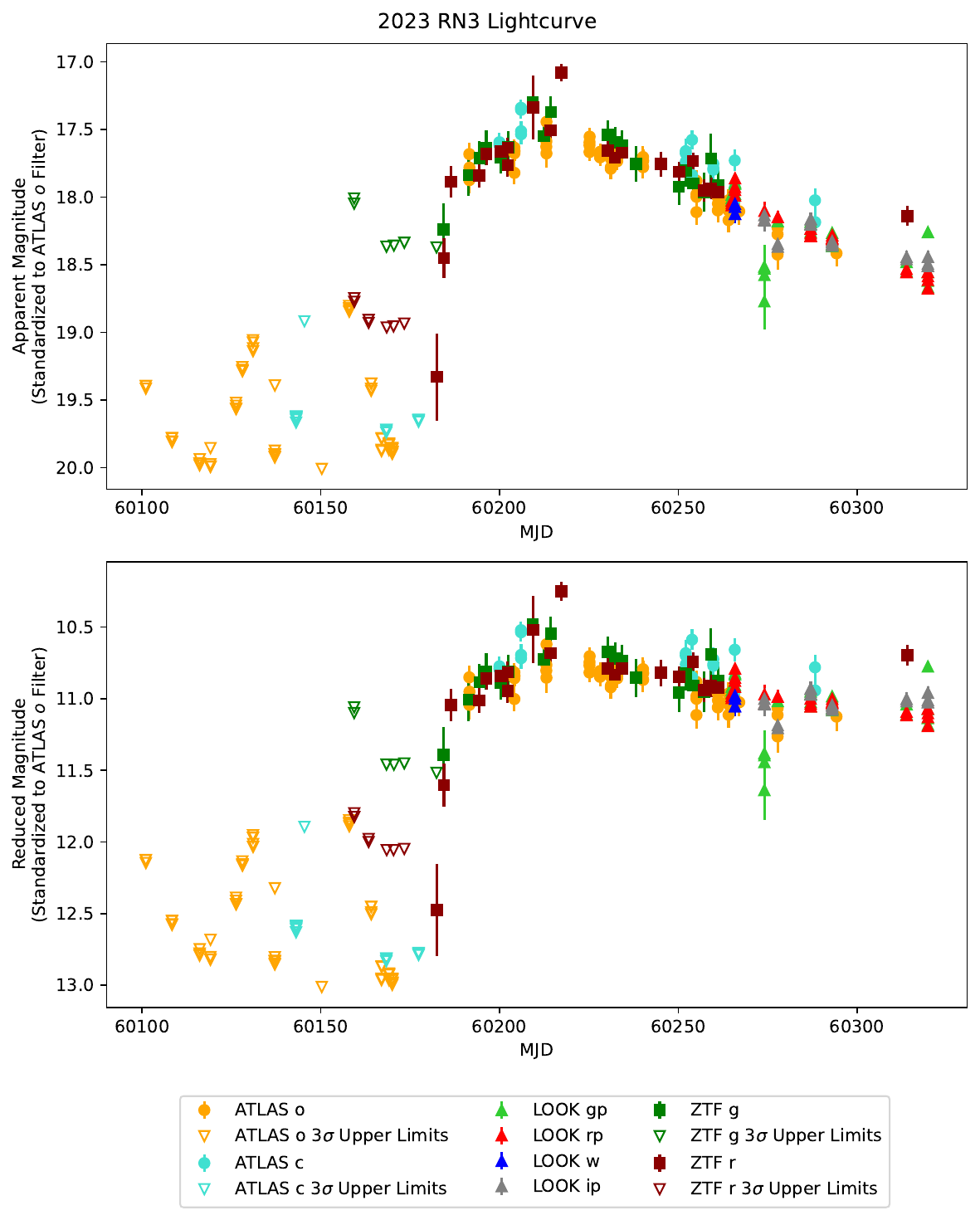}
\caption{Lightcurve of 2023 RN$_{3}$ plotting apparent magnitude (top) and reduced magnitude (bottom) across time. All magnitudes are corrected to the ATLAS $o$ filter wavelength. Filled markers indicate magnitude measurements of SNR $\geq3$, empty markers indicate $3\sigma$ upper limits.}
\label{2023RN3ApparentMagnitudeVsTimeCalibrated}
\end{figure}

To transform our magnitude measurements to the same wavelength, we select the ATLAS $o$ filter dataset as our standard filter on account of it having the largest number of observations and the longest time baseline, calculating and applying a color offset to the other datasets (listed in Table \ref{2023RN3ColourOffsets}). Magnitude offsets were calculated from the difference in median apparent magnitude between a given filter and the ATLAS $o$ filter across their overlapping timespan. 
For each color-corrected magnitude value, we query JPL Horizons for the corresponding geocentric and heliocentric distances of 2023 RN$_{3}$ at the time of observation and transform the apparent magnitudes of 2023 RN$_{3}$ into reduced magnitudes (the apparent magnitude scaled to geocentric and heliocentric distances of 1 au). 
Figure \ref{2023RN3ApparentMagnitudeVsTime} shows the apparent magnitude lightcurves of 2023 RN$_{3}$ in each filter, with the magnitudes standardized for the $o$-filter shown in Figure \ref{2023RN3ApparentMagnitudeVsTimeCalibrated}. The serendipitous datasets from ATLAS and ZTF allowed them to sample the rise of 2023 RN$_{3}$'s brightness, with ZTF making the first detection on 2023 August 26 and ATLAS making the last non-detection on 2023 August 21; targeted follow-up observations from LOOK were scheduled after the object was detected. Figures \ref{2023RN3ApparentMagnitudeVsTime} and \ref{2023RN3ApparentMagnitudeVsTimeCalibrated} show that over the course of 137 days, 2023 RN$_{3}$ suddenly increased in brightness at an initial rate of approximately 0.36 mag/day, reaching a peak apparent brightness of ${\sim}17.5$ mag, and has since levelled off to a constant magnitude after a slight decrease in brightness. 
This indicates that cometary activity may be continuing across this period and that 2023 RN3's increase in brightness is the onset of an epoch of continuous cometary activity. 
Extrapolating 2023 RN$_{3}$'s initial brightening rate to the $3\sigma$ limiting magnitude from \citet{2023RNAAS...7..263H} of 22.5 mag, which we assume to be the nuclear apparent magnitude of 2023 RN$_{3}$, we estimate the activity likely started within about 10 days of the first ZTF detection on 2023 August 26. 
\rcom{Our observed brightness maximum, combined with the limiting magnitude measurement from \citet{2023RNAAS...7..263H}, would imply a total increase in brightness of ${\Delta}m > 5.4$ mag, comparable to that observed for the large cometary outbursts of the Centaur 29P/Schwassman-Wachmann 1 \citep{2008A&A...485..599T,2016Icar..272..387M,2023PASJ...75..462L} and the comet 12P/Pons-Brooks \rcom{\citep[]{2023ATel16194....1M,2024LPICo3040.1363T}}}. 
From visual inspection of our reduced magnitude lightcurve, we do not see evidence of any secondary cometary outbursts by 2023 RN$_{3}$ with its brightness evolution consistent with a single sudden rise in brightness followed by a small decrease to a relatively constant magnitude. Our $3\sigma$ upper limits before 2023 August 26 preclude an epoch of cometary activity as bright as the observed 2023 epoch, however we cannot rule out lower-level cometary activity occurring before this date.

\begin{deluxetable*}{lr}[h] \label{2023RN3ColourOffsets}
\tablecaption{Magnitude offsets for 2023 RN$_{3}$ apparent magnitudes in each filter, correcting to the ATLAS $o$ filter.}
\tablecolumns{2}
\tablehead{
\colhead{Filter} & 
\colhead{Magnitude Offset}\\
\colhead{} & 
\colhead{(mag)} 
}
\startdata
ATLAS $c$&-0.563\\
ATLAS $o$&0.000\\
PS1 $g$ (ZTF)&-0.546\\
PS1 $r$ (ZTF)&0.063\\
PS1 $w$ (LOOK)&0.164\\
PS1 $g'$ (LOOK)&-0.593\\
PS1 $r'$ (LOOK)&-0.016\\
PS1 $i'$ (LOOK)&0.097\\
\enddata
\end{deluxetable*}

\subsection{Color Evolution} \label{ColourEvolution}

We use the dual filter coverage of ATLAS, ZTF, and LOOK to search for any changes in 2023 RN$_{3}$'s $c-o$ or $g-r$ color across time. As per Section \ref{BrightnessEvolution}, for each magnitude measurement we use the geocentric and heliocentric distances of 2023 RN$_{3}$ at the time of observation to calculate the corresponding reduced magnitude. We apply the two methods used in \citet{2024PSJ.....5..165D} for measuring color evolution for a varying lightcurve. 

The first method involves binning all the magnitude measurements from a given dataset by epoch of observation. ATLAS measurements that are separated in time by $<2$ days (its approximate observation cadence) are counted as part of an epoch. For ZTF and LOOK, there are often both $g$ and $r$ observations on the same night, thus we consider magnitude measurements separated by $<0.5$ days as part of one epoch. For each epoch, if observations in both filters are present, the mean reduced magnitude of each filter is calculated, and with its associated uncertainty calculated as the standard deviation of the magnitudes in each filter. The corresponding $c-o$ or $g-r$ color index of each is calculated by subtracting these mean magnitudes, with the associated uncertainties propagated in quadrature. 

The second method involves fitting polynomial splines to the magnitude values of a given filter for a given survey, allowing us to account for any sudden change in 2023 RN$_{3}$'s brightness across time. 
To ensure well-constrained polynomial fits, we only fit ZTF $g$ and $r$ observations before MJD 60270 due to the scarcity of data in these filters after these dates. Our ATLAS $c$ filter has too few measurements to sufficiently constrain the value this way, therefore we do not include it in our analysis.
\rcom{For each filter, we generate $10^3$ datasets of synthetic magnitudes. We create a synthetic magnitude dataset by duplicating the measured magnitude values and adding to each  magnitude value a random number generated from a Gaussian distribution centred on zero with a standard deviation equal to the measured magnitude's uncertainty. We then fit a 3rd order polynomial spline to each set of synthetic magnitude values. For a given magnitude measurement in a given filter, we calculate 2023 RN$_{3}$'s magnitude in the other filter at the same time of measurement as predicted by its fitted spline. These two magnitude measurements are then used to calculate 2023 RN$_{3}$'s ($g-r$) color at that time. This method is repeated for each of the $10^3$ synthetic datasets.}

\begin{figure}
\centering
\includegraphics[width=\textwidth]{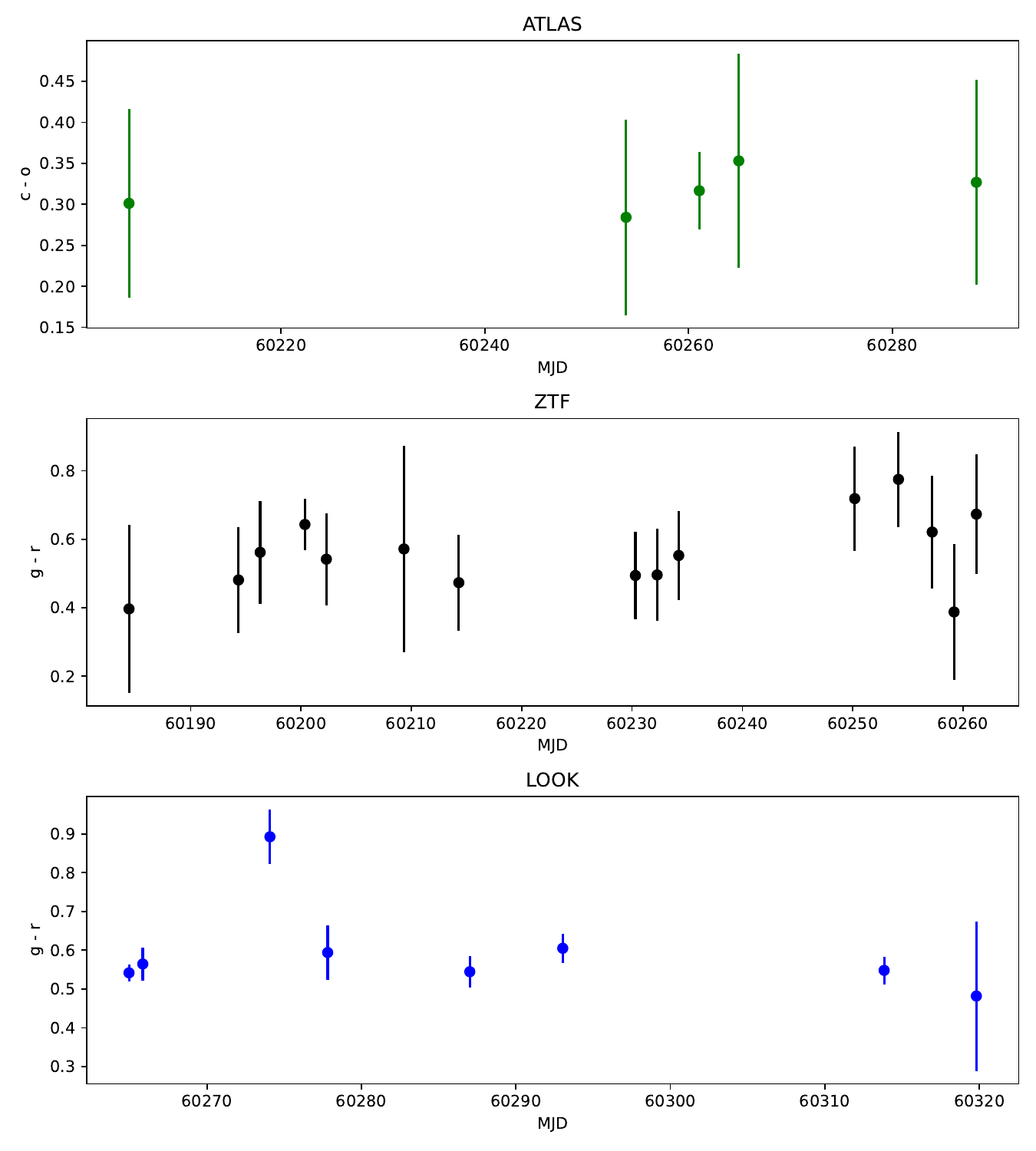}
\caption{ATLAS $c-o$ (upper), ZTF $g-r$ (center), and LOOK $g-r$ (lower) color index values calculated from binning observations by epoch plotted across time.}
\label{2023RN3ColourEvolution}
\end{figure}

\begin{figure}
\centering
\includegraphics[width=\textwidth]{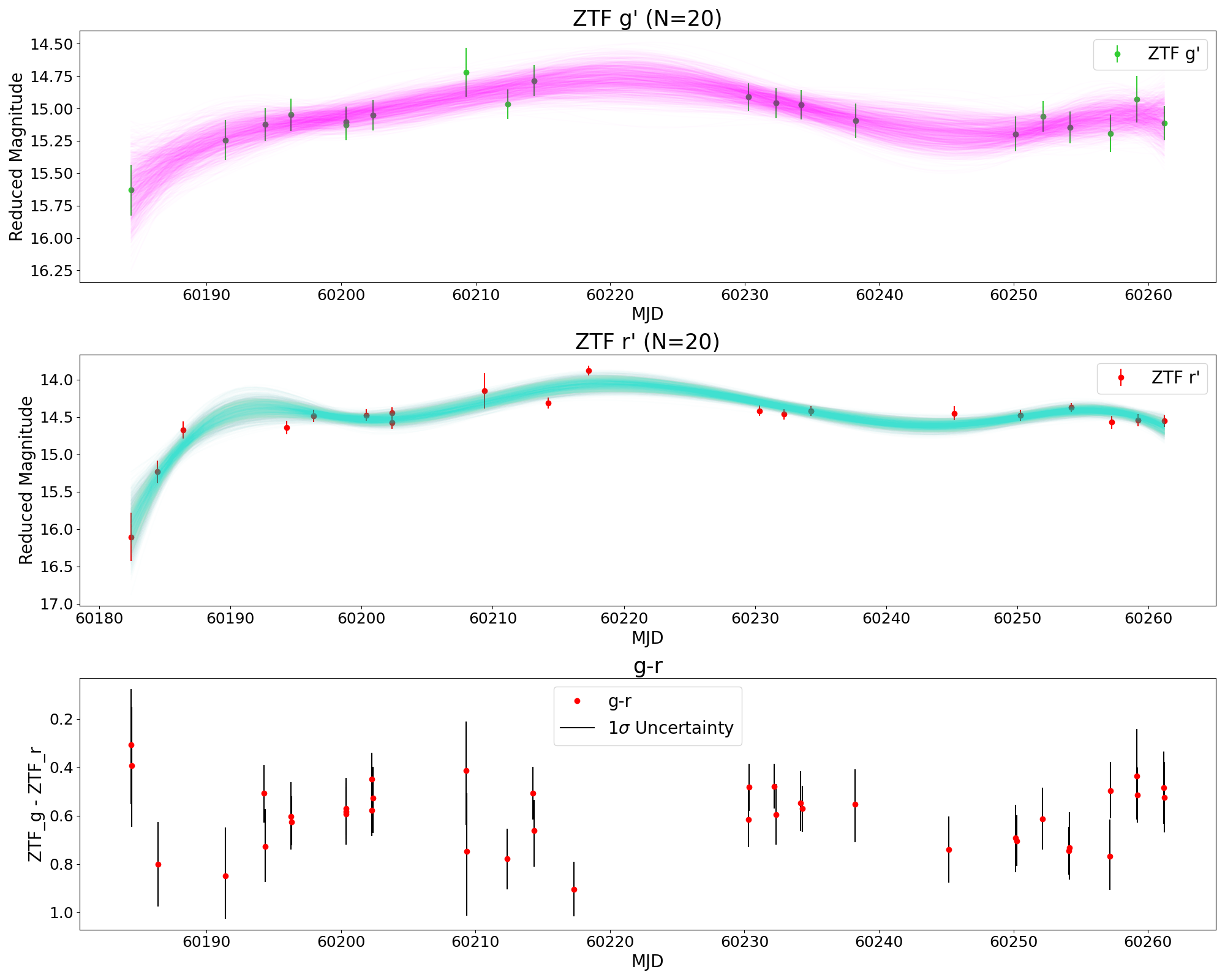}
\caption{Top: ZTF $g$ filter reduced magnitudes (green circles) vs. time with 3rd order interpolated splines fitted to synthetic data overplotted (green). Middle: ZTF $r$ filter reduced magnitudes (red circles) vs. time with 3rd order interpolated splines fitted to synthetic data overplotted (turquoise). Bottom: $g-r$ color index value calculated from the predicted magnitudes in each filter from the synthetic magnitudes plotted across time. Red circles denote nominal color index values with green lines highlighting the range that includes the central 68.3\% of synthetic values.}
\label{2023RN3ColourEvolutionLOOKSplines}
\end{figure}

\begin{figure}
\centering
\includegraphics[width=\textwidth]{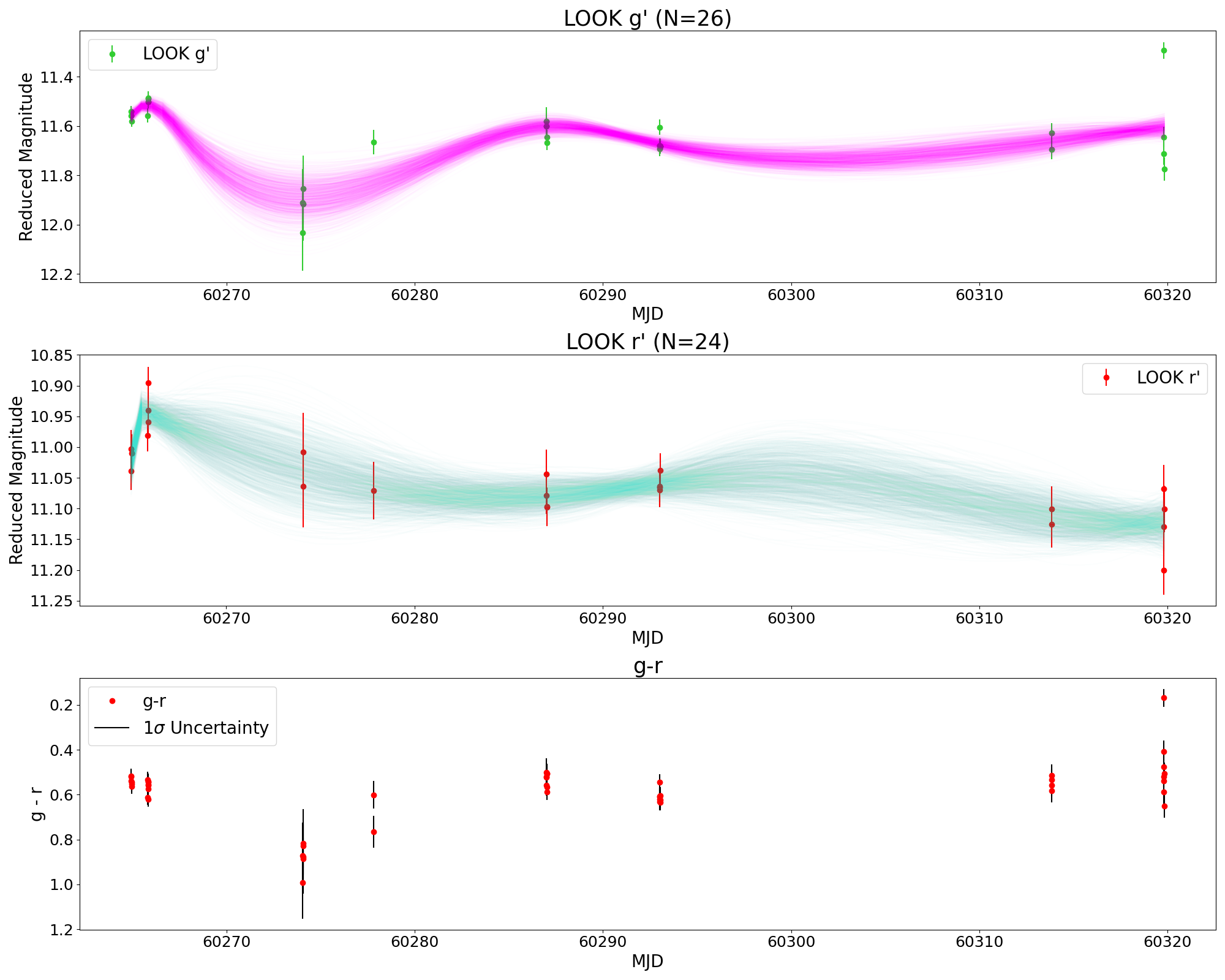}
\caption{Top: LOOK $g$ filter reduced magnitudes (green circles) vs. time with 3rd order interpolated splines fitted to synthetic data overplotted (green). Middle: LOOK $r$ filter reduced magnitudes (red circles) vs. time with 3rd order interpolated splines fitted to synthetic data overplotted (turquoise). Bottom: $g-r$ color index value calculated from the predicted magnitudes in each filter from the synthetic magnitudes plotted across time. Red circles denote nominal color index values with green lines highlighting the range that includes the central 68.3\% of synthetic values.}
\label{2023RN3ColourEvolutionZTFSplines}
\end{figure}

Figure \ref{2023RN3ColourEvolution} shows the resulting ZTF and LOOK color indices across time, calculated from binning data by epoch of observation. The ZTF and LOOK color indices calculated from the spline fitting method are shown in Figures \ref{2023RN3ColourEvolutionLOOKSplines} and \ref{2023RN3ColourEvolutionZTFSplines}, respectively. We see no significant change in color across our baseline of observation, with its $c-o$ and $g-r$ color indices remaining constant within uncertainties. The mean $c-o$ from ATLAS, calculated from spline fitting, is $0.316 \pm 0.023$ with a median of $0.317$. From binning our observations by epoch of observation, the mean $g-r$ values from ZTF and LOOK are $0.559 \pm 0.044$ and $0.597 \pm 0.029$ in their respective filter systems, with medians of $0.553$ and $0.556$ respectively.
When fitting our observations with splines, the mean $g-r$ values from ZTF and LOOK in their respective filter systems are $0.601 \pm 0.023$ and $0.592 \pm 0.008$, with medians of $0.583$ and $0.559$ respectively.
Our $g-r$ values for each filter system are consistent to $<2\sigma$ for both methods, and are also
consistent with the color measured by \citet{2023RNAAS...7..263H}, $g-r = 0.61 \pm 0.03$. For comparison, the corresponding solar color indices in these filter systems are $(c-o) = 0.29$ and $(g-r)_{PS1} = 0.39$ \citep[]{2018PASP..130f4505T,2018ApJS..236...47W}. Hence, assuming our data to be dominated by a coma as evidenced by both the comet's fuzzy appearance in previous observations \citep{2023RNAAS...7..263H} and the large increase in the comet's brightness, the color of 2023 RN$_{3}$'s coma is significantly redder than the Sun and similar to previously observed cometary dust comae \citep{2012Icar..218..571S}. 

\subsection{Surface Brightness Evolution} \label{PSFEvolution}

The increased spatial resolution of the 2023 RN$_{3}$ observations from LOOK and Liverpool telescope (0.39 and 0.30 $\prime\prime$/pixel, respectively) compared to ATLAS and ZTF (1.86 and 1.01 $\prime\prime$/pixel, respectively) allow us to analyze the structure of 2023 RN$_{3}$'s PSF and and investigate its evolution across time. We use our photometric detections of 2023 RN$_{3}$ where the comet's SNR is $\geq26$ to enable accurate analysis of its radial profile. We utilize Source Extractor Python \citep[]{1996A&AS..117..393B,2016JOSS....1...58B} to detect background field stars from the PanSTARRS Mean DR2 catalog \citep[]{2016arXiv161205560C,2020ApJS..251....3M,
2020ApJS..251....7F} and select the stars which satisfy the following criteria: 1) they are isolated sources i.e. no other detected source lies within 10 arcsec of their centroid to prevent contamination; 2) the difference between their PanSTARRS Mean DR2 PSF magnitudes and Kron magnitudes in the $i$ filter is less than 0.05 mag to separate stars from galaxies; and 3) their $g$-filter PanSTARRS Mean DR2 PSF magnitude is brighter than 20th mag to ensure they are of comparable brightness to 2023 RN$_{3}$. We then select the 10 stars whose peak fluxes as measured with Source Extractor Python are closest to that of 2023 RN$_{3}$ to compare to the comet. We use the \textit{RadialProfile} function of the \textit{photutils} Python library \citep{larry_bradley_2023_7946442} to measure the radial profile of 2023 RN$_{3}$ in comparison to the background stars, utilizing the background flux value calculated using the \textit{Background} function from \textit{Source Extractor Python} to remove the background noise contribution from each radial profile.

\begin{figure}
\centering
\includegraphics[height=0.85\textheight]{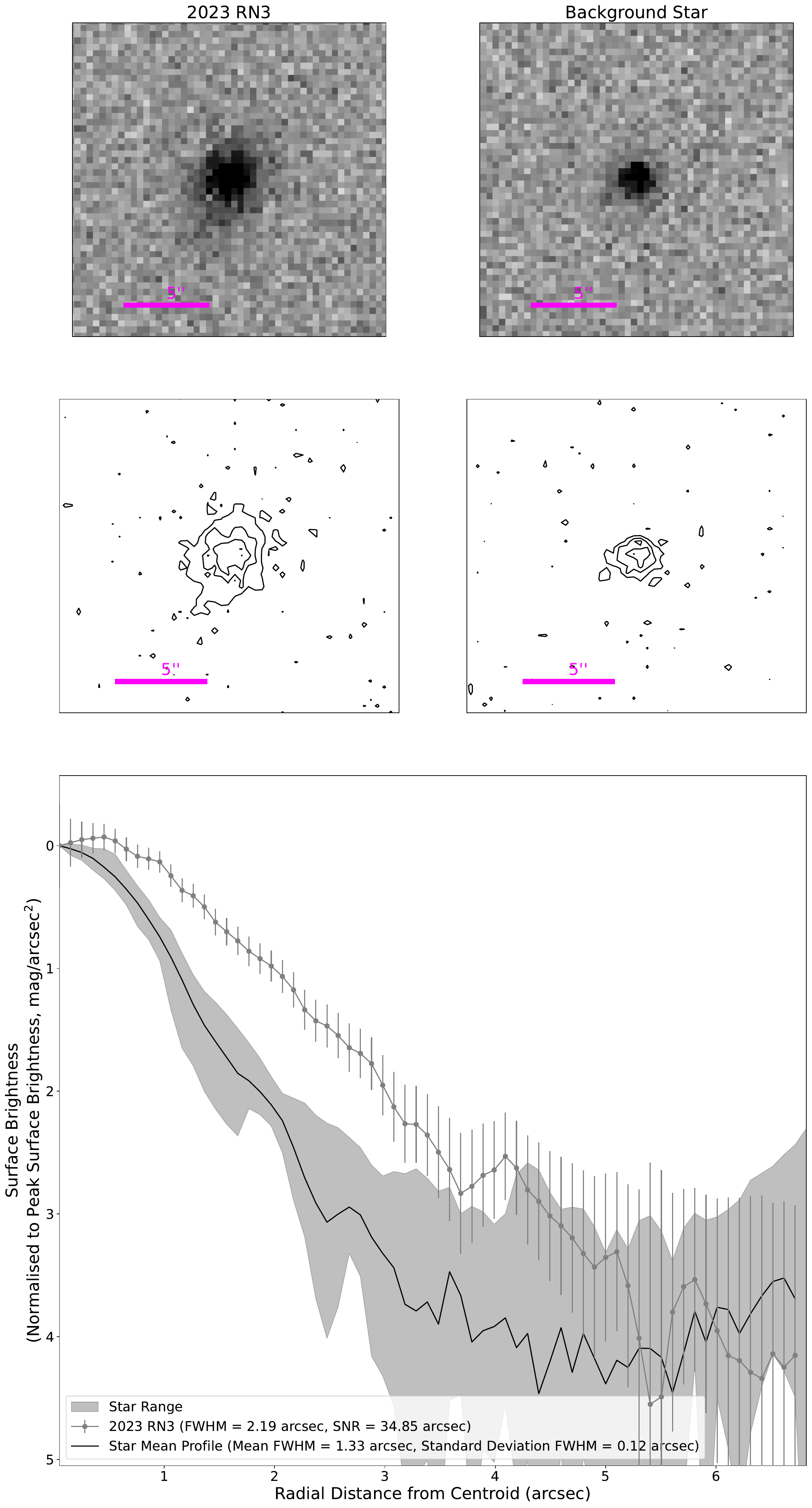}
\caption{Top row: 2024 January 04 (MJD 60313.8750614) i-filter LOOK observation of 2023 RN$_{3}$ (left) and a nearby background star (right). Center row: Corresponding contour plots of the PSFs of 2023 RN$_{3}$ (left) and the same background star (right). Bottom: Radial profile of 2023 RN$_{3}$ (grey circles) compared to mean radial profile (black line) and range of profiles (grey shaded) of 10 nearby field stars from 2024 January 04 i-filter LOOK observation, plotting instrumental surface brightness (in units of magnitudes/arcsec$^{2}$) normalised to the peak brightness of 2023 RN$_{3}$ vs. radial distance from the PSF centroid. Note: this figure comprises part of a figure set which can be found in the online published version of this paper.}
\label{2023RN3LOOKContourPlots}
\end{figure}

\begin{figure}
\centering
\includegraphics[height=0.85\textheight]{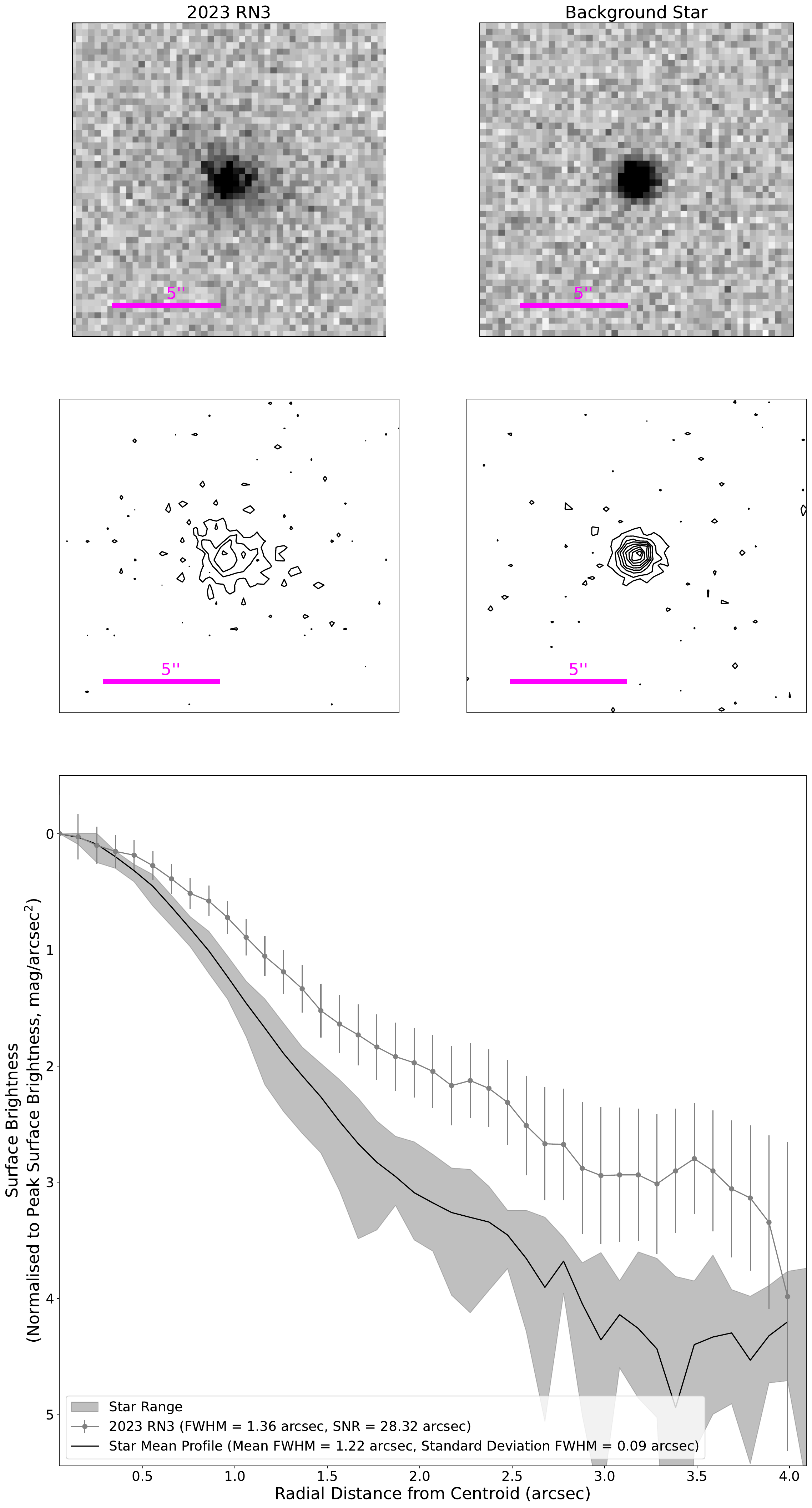}
\caption{Top row: 2023 December 14 (MJD 60292.826273) SDSS-I filter Liverpool Telescope observation of 2023 RN$_{3}$ (left) and a nearby background star (right). Center row: Corresponding contour plots of the PSFs of 2023 RN$_{3}$ (left) and the same background star (right). Bottom: Radial profile of 2023 RN$_{3}$ (grey circles) compared to mean radial profile (black line) and range of profiles (grey shaded) of 10 nearby field stars from 2023 December 14 SDSS-I filter LOOK observation, plotting instrumental surface brightness (in units of magnitudes/arcsec$^{2}$) normalised to the peak brightness of 2023 RN$_{3}$ vs. radial distance from the PSF centroid. Note: this figure comprises part of a figure set which can be found in the online published version of this paper.}
\label{2023RN3LiverpoolTelescopeContourPlots2023-12-14MJD60292.826273}
\end{figure}

Figures \ref{2023RN3LOOKContourPlots} and \ref{2023RN3LiverpoolTelescopeContourPlots2023-12-14MJD60292.826273} show respectively LOOK and Liverpool Telescope observations of 2023 RN$_{3}$ and for comparison a field star with similar peak flux, their corresponding contour plots, and the radial profile of 2023 RN$_{3}$ as compared to the 10 field stars of similar brightness, with the linear fit to the profile of 2023 RN$_{3}$ overplotted. We calculate the full width half maximum (FWHM) of 2023 RN$_{3}$ and the comparison stars using the equation:
\begin{equation}
    FWHM = 2\sqrt{ln(2)(a^{2}+b^{2})}
\end{equation}
where $a$ and $b$ are the semimajor and semiminor axes of the fitted PSF to a given source detected by Source Extractor Python.
The FWHM of 2023 RN$_{3}$ and the mean FWHM of the 10 comparison stars for each LOOK and Liverpool Telescope observation are included in the caption of each corresponding figure in Figure Sets 1 and 2. We find that for each observation, the FWHM of 2023 RN$_{3}$ is extended compared to that of the background field stars and/or the radial profile of 2023 RN$_{3}$ deviates from the background star profile. This indicates the presence of a spatially extended coma, consistent with the findings of \citet{2023RNAAS...7..263H}.

\subsection{$Af\rho$}
To examine 2023 RN$_{3}$'s dust production rate, we use the parameter $Af\rho$, introduced by \citet{1984AJ.....89..579A}, which is proportional to the mass loss rate of an observed comet. The product of the cometary albedo $A$, dust grain filling factor $f$, and aperture radius $\rho$, $Af\rho$ can be calculated via:

\begin{equation}
    Af\rho = \frac{4R_{H}^{2}\Delta^{2}}{\rho}\frac{F_{com}}{F_{Sun}}
\end{equation}
where $\Delta$ is the geocentric distance of an observed comet (measured in centimetres), $R_{H}$ is its heliocentric distance (measured in au), $\rho$ the aperture radius used to measure its brightness (measured in cm), $F_{com}$ is the measured flux of the comet, and $F_{Sun}$ is the solar flux measured at 1 au. This equation can be alternatively expressed in terms of the magnitudes of the comet $M_{com}$ and Sun $M_{Sun}$ as:

\begin{equation}
    Af\rho = \frac{4R_{H}^{2}\Delta^{2}}{\rho}10^{0.4(M_{Sun}-M_{com})}
\end{equation}
We use an aperture radius of $\rho = 20,000$ km as per Section \ref{BrightnessEvolution}, and the same geocentric and heliocentric distances as used in Section \ref{ColourEvolution}. The cometary activity levels exhibited by 2023 RN$_{3}$ are wavelength dependent, as the color of the material emitted by the comet relative to the Sun will have a corresponding effect on the measured $Af\rho$ values.
As 2023 RN$_{3}$ has not undergone any significant color evolution across our dataset, we take our lightcurve corrected to the ATLAS $o$ filter from Figure \ref{2023RN3ApparentMagnitudeVsTimeCalibrated} in Section \ref{BrightnessEvolution} to calculate $Af\rho$. For this calculation, we apply the ATLAS $o$ filter solar magnitude measurement of $-26.982$ \citep{2024PSJ.....5...25G}.
$Af\rho$ is also dependent on the solar phase angle of the comet at the time of measurement due to the light-scattering properties of the coma dust particles. We correct our $Af\rho$ values for effects of phase angle using the Schleicher-Marcus\footnote{\url{https://asteroid.lowell.edu/comet/dustphase/details}} cometary phase function \citep[]{1998Icar..132..397S,2007ICQ....29...39M,Schleicher_et_al_2010}. 

\begin{figure}
\centering
\includegraphics[width=\textwidth]{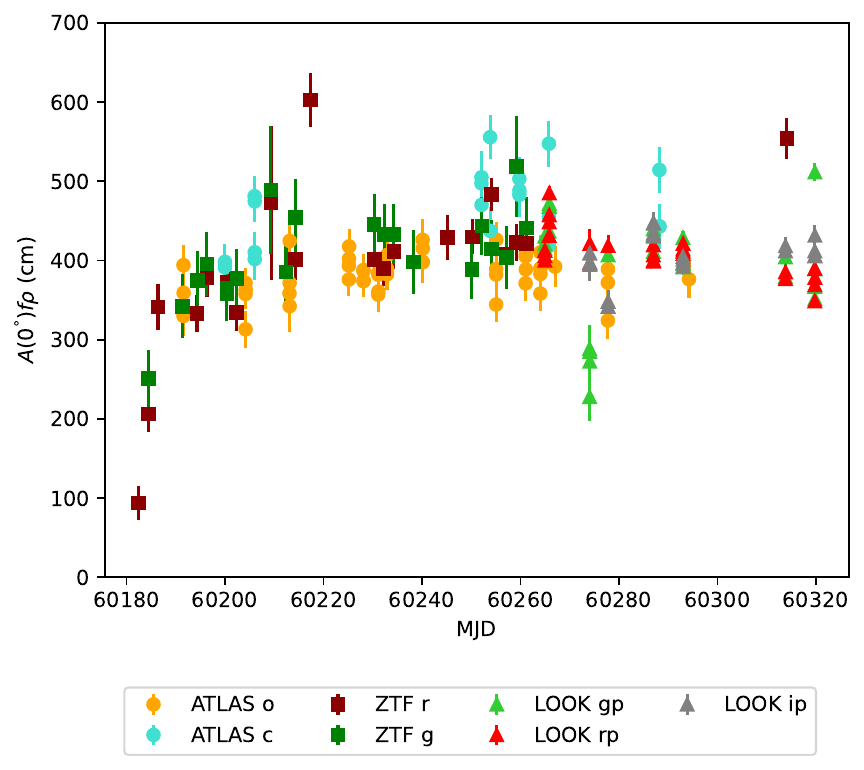}
\caption{$A(0\degree)f\rho$ values of 2023 RN$_{3}$ across time. $A(0\degree)f\rho$ values have been corrected for color and phase effects.}
\label{2023RN3Afrho}
\end{figure}

The resulting phase-corrected values, $A(0\degree)f\rho$, are plotted across time in Figure \ref{2023RN3Afrho}, showing 2023 RN$_{3}$ to rapidly increase in activity before plateauing around a maximum $A(0\degree)f\rho$ value of approximately $400$ cm for most of our observations. This is consistent with continuous and sustained activity instead of a brief outburst. We consider the continuous nature of 2023 RN$_{3}$'s cometary activity sufficient to permit comparison of its $Af\rho$ value to other objects. \rcom{2023 RN$_{3}$'s maximum $A(0\degree)f\rho$ value is greater than that of most JFCs \citep[]{2023Icar..39115340S,2024PSJ.....5...25G,2024arXiv240408618K} yet is far less than those of outbursts exhibited by the JFC 17P/Holmes  \rcom{\citep{2008LPI....39.1627T,2009AJ....138..625L,2010MNRAS.409.1682T}} and the Centaur 29P Schwassman-Wachmann 1 \rcom{\citep[]{2010MNRAS.409.1682T,2023PASJ...75..462L}}, which increased the apparent brightness of the respective objects by several magnitudes, and whose maximum $Af\rho$ values were of the order of }\rcom{$10^3 - 10^5$ cm. This difference in $Af\rho$ values may be due to different sizes or numbers of active regions on the objects' surfaces, though a comparison of such active regions is beyond the scope of this paper.} 

\subsection{Orbital Evolution}
\begin{figure}
\centering
\includegraphics[width=\textwidth]{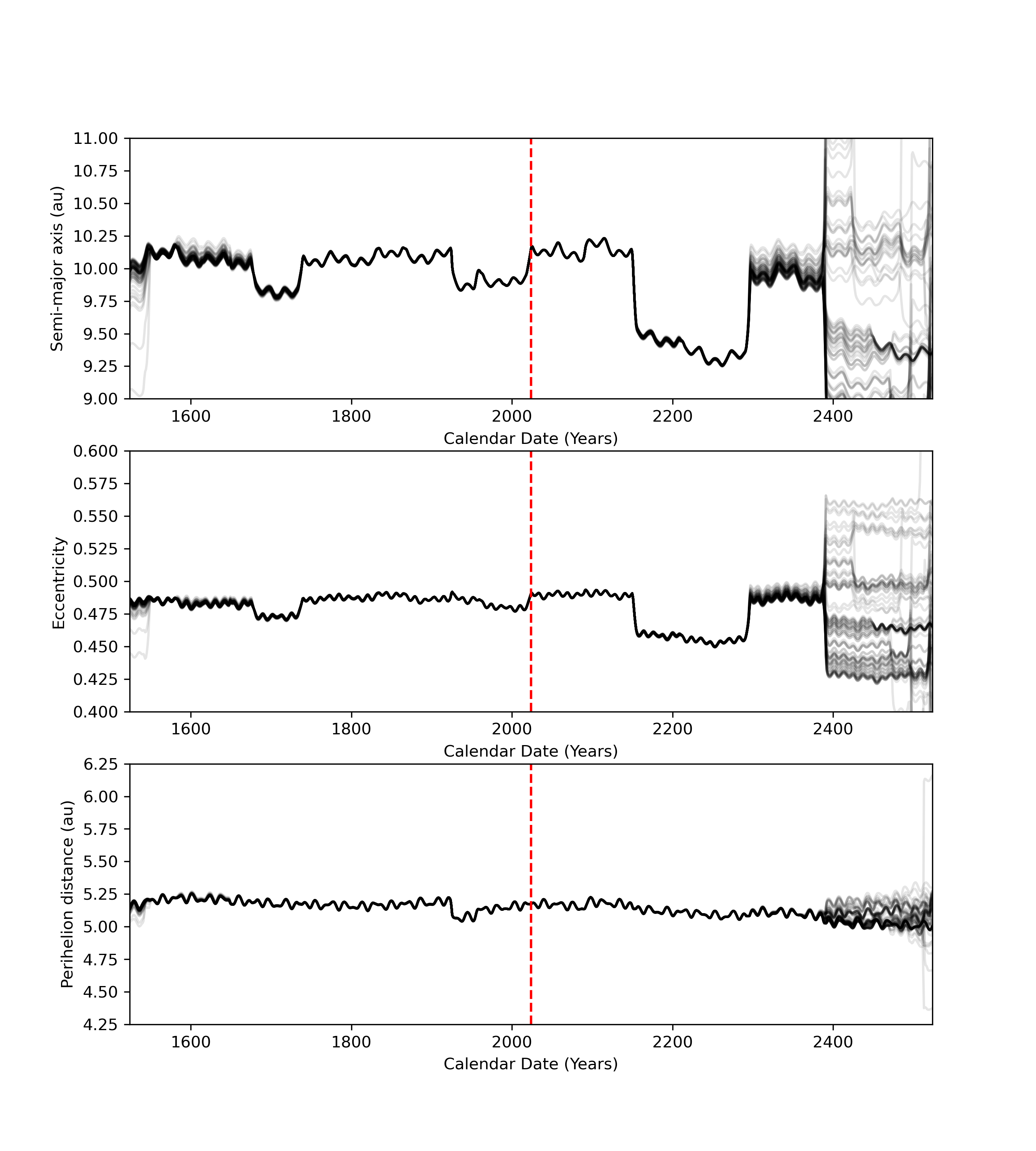}
\caption{Orbital evolution calculations of the 2023RN$_3$ nominal orbit and 100 clones. Top plot; semi-major axis $a$ versus date. The red dashed line indicates the starting epoch of the orbital integrations. The y-axis range of the top plot has been chosen to clearly show the small jumps in $a$ near the current epoch. Middle and bottom plots; as for the top plot but showing eccentricity $e$ and perihelion distance $q$.  }
\label{2023RN3orbit}
\end{figure}

As mentioned above, \cite{2024ApJ...960L...8L} found evidence that cometary activity in Centaurs and high-perihelion JFCs may be triggered by rapid changes in semi-major axis due to planetary perturbations. To investigate this possibility for 2023 RN$_{3}$, we  performed an N-body simulation using \textit{REBOUND} \citep{Rein2012} and \textit{astropy} \citep[]{2013A&A...558A..33A,2018AJ....156..123A,2022ApJ...935..167A}. We obtained the most recent orbital elements with epoch JD 2460222.5 $\equiv$  2023.759 TDB from the JPL solar system Database via the JPL SSDB API, calculated from observations up to 13 January 2024, and generated an additional 100 clones using the associated covariance matrix.  Including the 8 major planets, we integrated all bodies for $\pm 500$ years from the orbital epoch using the {\sc ias15} integrator \citep{Rein2015}. The resulting orbital evolution of all 101 test objects are shown in Figure \ref{2023RN3orbit}. We found that the orbit of 2023 RN$_{3}$ is well constrained between the years 1549 and 2388.  Planetary encounters on those dates  significantly increase orbital uncertainties outside this date range. At the current epoch, 2023 RN$_{3}$ has just experienced a rapid increase in semi-major axis of $\Delta a=+0.3$ au due to a distant encounter with Jupiter, with a small simultaneous increase in eccentricity $e$ maintaining the same perihelion distance $q$.   We note that \cite{2024ApJ...960L...8L} found almost all activity in Centaurs correlated with rapid {\em decreases} in $a$, with only 4 out of 56 objects having undergone increases in $a$.  The most similar object in their study was C/2013 P4 (Pan-STARRS), which also underwent $\Delta a=+0.3$ au 176 years ago, although of course we have no information on its activity at that time. Hence it is unclear whether this small $3$\%  increase in $a$ would or could lead to the sudden onset of significant activity observed by us in 2023 RN$_{3}$.

\section{Summary and Conclusions} \label{Discussion}

We analyzed the 2023 cometary activity of the JFC 2023 RN$_{3}$ using serendipitous observations from the ATLAS and ZTF surveys combined with targeted follow-up observations from LOOK and the Liverpool Telescope\rcom{. This resulted in a lightcurve spanning seven months of high-cadence observations. }\rcom{We used this lightcurve of 2023 RN$_{3}$ to analyze its evolution in brightness, color index, and activity levels. We also utilized our higher-resolution observations from LOOK and the Liverpool Telescope to analyze the radial profile of 2023 RN$_{3}$. }Our key findings are:

\begin{itemize}
    \item The lightcurve of 2023 RN$_{3}$ is indicative of the onset of an epoch of continuous cometary activity, commencing about 10 days before its first detection on 2023 August 26, and reaching a maximum brightness of about $17.5$ mag, corresponding to a brightness increase of ${\geq}5.4$ mag based on upper limits from \citet{2023RNAAS...7..263H}. 
    \item We do not observe any significant change in color across our observations.
    \item Analysis of 2023 RN$_{3}$'s surface brightness radial profile shows significant extension compared to background stars, indicating the presence of a spatially-extended coma, consistent with \citet{2023RNAAS...7..263H}. 
    \item We measure a maximum dust production rate of $A(0^{\degree})f\rho \sim 400$ cm, larger than that of most JFCs, but smaller than that observed for the outbursts of 29P and 17P. 
    \item We find that 2023 RN$_{3}$ has experienced an increase in semimajor axis of $a = +0.3$ au due to a planetary encounter with Jupiter, coinciding with its current active epoch.  
\end{itemize}

With a heliocentric distance at the time of its detected onset of cometary activity of 5.301 au, 2023 RN$_{3}$ resides beyond the so-called `ice line' at ${\sim}3$ au, interior to which direct sublimation of water ice is the dominant mechanism for observed cometary activity \citep{2004come.book..317M,2017PASP..129c1001W}. This rules out direct sublimation of water-ice as the mechanism behind 2023 RN$_{3}$'s 2023 cometary activity. \rcom{Possible drivers of this cometary activity could be the volatile species of CO or CO$_{2}$ which are thought to dominate observed cometary activity beyond the `ice line' \citep[]{2012ApJ...758...29A,2012ApJ...752...15O,2013Icar..226..777R,2015SSRv..197....9C,2015ApJ...814...85B,2017PASP..129c1001W}.}\rcom{ While we have shown that 2023 RN$_{3}$ has recently experienced a slight increase in semimajor axis due to a planetary encounter, whether this has led to the observed onset of cometary activity remains ambiguous as cometary activity in Centaurs and JFCs triggered by planetary encounters is normally associated with rapid decreases in the objects' semimajor axes \citep{2024ApJ...960L...8L}. }\rcom{The lack of any detected periodic outbursts by 2023 RN$_{3}$ may be due to its nucleus having a bilobate structure akin to that of 29P/Schwassman-Wachmann 1 \citep{2024NatAs.tmp..153F}, shielding a possible patch of volatiles from direct insolation until the onset of its recent active epoch. Confirmation of the nuclear shape of 2023 RN$_{3}$ via lightcurve analysis or direct imaging could help test this theory.}

Estimates of the maximum dust-production levels across 2023 RN$_{3}$'s 2023 cometary activity are considerably smaller than those of cometary outbursts with similar changes in brightness. 2023 RN$_{3}$'s $Af\rho$ maximum is consistent with the JFC population as a whole \citep[]{2023Icar..39115340S,2024PSJ.....5...25G,2024arXiv240408618K}, albeit at the high end of the distribution, 
and is also consistent with those exhibited by the Centaur population (ranging from ${\sim}10^{2}-10^{5}$ cm). 
This further highlights 2023 RN$_{3}$'s relevance to the study of cometary activity in both these populations.

We highlight that this object, and its corresponding cometary activity, was detected via serendipitous observations from ATLAS. The upcoming Legacy Survey of Space and Time by the Vera C. Rubin Observatory \citep[]{2009arXiv0912.0201L,2019arXiv190108549J,2019ApJ...873..111I,2023ApJS..266...22S}, with its high observation cadence and faint limiting magnitude, will allow us to track the continued brightness evolution of 2023 RN$_{3}$ and detect any changes in activity such as future cometary outbursts. LSST's faint limiting magnitude of ${\sim}24$ mag \citep{2022ApJS..258....1B} will allow us to discover many more such instances of cometary activity across the solar system, helping us to further shed light on the evolution of short period comets.

\begin{acknowledgements}

\section{Acknowledgements}

MMD was supported by the UK Science Technology Facilities Council (STFC) grant ST/V506990/1.
MES was supported by the UK Science Technology Facilities Council (STFC) grant ST/X001253/1. 
AF was supported by UK STFC grant ST/X001253/1.
MSK was supported by the NASA solar system Observations program (80NSSC20K0673).
LJS acknowledges support by the European Research Council (ERC) under the European Union’s Horizon 2020 research and innovation program (ERC Advanced Grant KILONOVA No. 885281).
JM acknowledges support from the Department for the Economy (DfE) Northern Ireland postgraduate studentship scheme. 
This research has made use of data and/or services provided by the International Astronomical Union's Minor Planet Center.
This research has made use of services provided by NASA's Astrophysics Data System. 
This work has made use of data and services provided by the Horizons system of the Jet Propulsion Laboratory. 

\rcom{This work has made use of data from the Asteroid Terrestrial-impact Last Alert System (ATLAS) project. The Asteroid Terrestrial-impact Last Alert System (ATLAS) project is primarily funded to search for near earth asteroids through NASA grants NN12AR55G, 80NSSC18K0284, and 80NSSC18K1575; byproducts of the NEO search include images and catalogs from the survey area. This work was partially funded by Kepler/K2 grant J1944/80NSSC19K0112 and HST GO-15889, and STFC grants ST/T000198/1 and ST/S006109/1. The ATLAS science products have been made possible through the contributions of the University of Hawaii Institute for Astronomy, the Queen?s University Belfast, the Space Telescope Science Institute, the South African Astronomical Observatory, and The Millennium Institute of Astrophysics (MAS), Chile.}

This work is based on observations obtained with the Samuel Oschin Telescope 48 inch Telescope at the Palomar Observatory as part of the Zwicky Transient Facility project. Major funding has been provided by the U.S. National Science Foundation under grant No. AST-1440341 and by the ZTF partner institutions: the California Institute of Technology, the Oskar Klein Centre, the Weizmann Institute of Science, the University of Maryland, the University of Washington, Deutsches Elektronen-Synchrotron, the University of Wisconsin-Milwaukee, and the TANGO Program of the University System of Taiwan.

This work makes use of observations from the Las Cumbres Observatory global telescope network.  Observations with the LCOGT 1m were obtained as part of the LCO Outbursting Objects Key (LOOK) Project (KEY2020B-009).

This work has utilized observations from the Liverpool Telescope. The Liverpool Telescope is operated on the island of La Palma by Liverpool John Moores University in the Spanish Observatorio del Roque de los Muchachos of the Instituto de Astrofisica de Canarias with financial support from the UK Science and Technology Facilities Council.

\end{acknowledgements}

Data Access: 
All photometric data used in this study is provided in full as supplementary information accompanying this paper.  Raw and Calibrated Observations from Las Cumbres Observatory used in this study are available at the LCO Science Archive (\url{https://archive.lco.global}; proposal code KEY2020B-009) after an embargo/proprietary period of 12 months. Calibrated observations from the Liverpool Telescope used in this study are available at the Liverpool Telescope Data Archive (\url{https://telescope.ljmu.ac.uk/cgi-bin/lt_search}; observing program XPQ23B01) after a proprietary period of one year after the end of the observing semester when the observations were acquired (observing semester 2023B).

\facilities{ATLAS (Chile, Haleakala, Mauna Loa, and South Africa telescopes), PO:1.2m (ZTF), LCOGT (1-m telescopes), Liverpool:2m}

\software{Astropy 
\citep[]{2013A&A...558A..33A,2018AJ....156..123A,2022ApJ...935..167A},
BANZAI \citep{2018SPIE10707E..0KM},
\calviacat \citep{2021zndo...5061298K},
{\sc ias15} \citep{Rein2015},
Jupyter Notebook \citep{soton403913},
math \citep{van1995python},
Matplotlib \citep{Hunter:2007},
NEOExchange \citep{2021Icar..36414387L},
Numpy \citep{2011CSE....13b..22V,harris2020array},
os \citep{van1995python},
Pandas \citep{reback2020pandas},
Photutils \citep{larry_bradley_2023_7946442},
python (\url{https://www.python.org}),
REBOUND \citep{Rein2012},
SAOImageDS9 \citep{2019zndo...2530958J},
SciPy \citep{2020NatMe..17..261V},
Source Extractor Python \citep[]{1996A&AS..117..393B,2016JOSS....1...58B}}

\bibliography{bibfile-2.bib}{}
\bibliographystyle{aasjournal}

\end{document}